\documentclass[a4paper]{article}

\usepackage[pages=all, color=black, position={current page.south}, placement=bottom, scale=1, opacity=1, vshift=5mm]{background}
\SetBgContents{
	\tt    
}      

\usepackage[margin=1in]{geometry} 

\usepackage{amsmath}
\usepackage{amsthm}
\usepackage{amssymb}
\usepackage{subfig}
\usepackage{graphicx}

\usepackage[utf8]{inputenc}
\usepackage{hyperref}
\hypersetup{
	unicode,
	pdfauthor={Author One, Author Two, Author Three},
	pdftitle={A simple article template},
	pdfsubject={A simple article template},
	pdfkeywords={article, template, simple},
	pdfproducer={LaTeX},
	pdfcreator={pdflatex}
}

\newcommand{\be}{\begin{equation}}
\newcommand{\ee}{\end{equation}}
\newcommand{\bea}{\begin{eqnarray}}
\newcommand{\eea}{\end{eqnarray}}

\usepackage[sort&compress,numbers,square]{natbib}

\theoremstyle{plain}

\theoremstyle{definition}

\usepackage{graphicx, color}
\graphicspath{{fig/}}

\usepackage{algorithm, algpseudocode} 
\usepackage{mathrsfs} 

\usepackage{lipsum} 

\begin{document}
\title{Scalar-Field Singularities and Cyclic Cosmology in Brans–Dicke Theory}

\author{Konrad Marosek${^1} {^{\divideontimes}}$ \and Adam Balcerzak$^{2,3}$}

\date{
\small{$^1$Institute of Mathematics, Physics and Chemistry, Maritime University of Szczecin, Wa{\l }y Chrobrego 1-2, 70-500 Szczecin, Poland \\
$^2$Institute of Physics, University of Szczecin,  Wielkopolska 15, 70-451 Szczecin,  Poland \\
$^3$Copernicus Center for Interdisciplinary Studies, Szczepa\'nska 1/5, 31-011 Krak\'ow, Poland \\
$^{\divideontimes}$Corespondence: k.marosek@pm.szczecin.pl}}

\maketitle

\begin{abstract}
We investigate a Brans–Dicke cosmological model with a scalar-field potential and derive an analytical solution for a power-law ansatz for the scalar field. The resulting cosmological evolution terminates in a scalar-field singularity at which the scalar field diverges and the effective gravitational coupling tends to zero, while the scale factor remains finite and the energy density and pressure diverge. We then explore a heuristic interpretation of this behaviour in which the rapid decrease of the effective gravitational coupling near the singularity could be associated with accelerated black-hole evaporation. We phenomenologically postulate a transition to a subsequent cosmological branch, although a smooth transition preserving the same scalar-field potential on both branches has not been obtained within the ansatz considered. The physical mechanism underlying such a transition therefore remains unknown. Finally, we discuss possible, highly speculative connections between this scenario and the Little Red Dots observed by the James Webb Space Telescope. These connections are not predictions of the present model but motivate further investigation of the possible implications of scalar-field singularities for cyclic cosmological scenarios.
\end{abstract}

\section{Introduction}
\label{sec:intro}

One of the fundamental challenges of general relativity is the occurrence of spacetime singularities. In the classical description, singular behavior is closely associated with geodesic incompleteness, which was established in the Hawking--Penrose singularity theorems as a key indicator of the breakdown of the spacetime description \cite{HawkingElis}. In particular, a spacetime is considered geodesically incomplete if at least one timelike or null geodesic cannot be extended to arbitrary values of its affine parameter. Following the discovery of the accelerated expansion of the Universe \cite{Riess,Perlmutter,Tonry,Knop,Tegmark} and the possibility of phantom dark energy, characterized by an equation-of-state parameter $w<-1$ \cite{Hinshaw}, a number of exotic scenarios describing the future evolution of the Universe have been proposed \cite{Haro}. These developments have also demonstrated that geodesic incompleteness alone is insufficient to characterize the physical nature and strength of a singularity.

Among the finite-time singularities occurring at $t\rightarrow t_s$ are the Big-Rip singularity (Type I) \cite{Dabrowski1,Caldwell}, the Sudden Future Singularity (Type II) \cite{Barrow,Barrow1}, the Finite Scale Factor Singularity (Type III) \cite{Odintsov1,Dabrowski2}, the Big-Separation Singularity (Type IV) \cite{Odintsov2}, and the $(w)$-singularity (Type V) \cite{Dabrowski3,Fernandez}. Other exotic scenarios arise only asymptotically as $t\rightarrow\infty$, including the Little Rip \cite{Frampton} and Pseudo-Rip \cite{Frampton1} scenarios. These different types of singular behavior are distinguished by the asymptotic properties of the scale factor $a$, the energy density $\rho c^2$, the pressure $p$, and, in the case of a Type V singularity, the barotropic equation-of-state parameter $w$.

The occurrence and properties of these singularities are closely related to the energy conditions imposed on the matter content of the Universe \cite{Curiel}. Depending on the particular scenario, one or more of the standard energy conditions may be violated. For a perfect fluid, the null energy condition (NEC), weak energy condition (WEC), strong energy condition (SEC), and dominant energy condition (DEC) can be written as
\bea
\text{(NEC)}
\qquad
\rho c^2+p\geq0, \notag
\eea
\bea
\text{(WEC)}
\qquad
\rho c^2\geq0,
\qquad
\rho c^2+p\geq0, \notag
\eea
\bea
\text{(SEC)}
\qquad
\rho c^2+p\geq0,
\qquad
\rho c^2+3p\geq0, \notag
\eea
and
\bea
\text{(DEC)}
\qquad
\rho c^2\geq0, \notag
\qquad
-\rho c^2\leq p\leq\rho c^2.
\eea
Violations of these conditions are characteristic of many exotic cosmological scenarios, particularly those involving phantom-like matter with $w<-1$. The behavior of geodesics near different types of singularities can also vary substantially. Their strength can be further characterized by means of the criteria introduced by Tipler \cite{Tipler} and Królak \cite{Królak}, which provide a more refined description than geodesic incompleteness alone.

Several modified-gravity and semiclassical scenarios have been proposed in which an initially strong singularity can be weakened or avoided altogether. Examples include $f(R,T)$ gravity \cite{Houndjo}, models with varying fundamental constants \cite{Marosek}, and scenarios in which Hawking radiation from a cosmological horizon is taken into account before the singularity is reached \cite{Nojiri,Elizalde}. These approaches indicate that the nature of a future singularity can depend sensitively on the underlying gravitational dynamics and on the physical assumptions concerning the matter and gravitational sectors.

Among the modified-gravity frameworks capable of altering the behavior of singularities, scalar-tensor theories are particularly interesting. A prominent example is Brans--Dicke gravity \cite{BransDicke}, which provides a scalar-tensor extension of general relativity. In this framework, the scalar field $\phi$ is dynamical and may depend on both time and spatial coordinates. In the Jordan frame, the scalar field is nonminimally coupled to the spacetime geometry and is accompanied by a kinetic term. Its dynamical evolution provides an effective description of variations in the gravitational coupling, which can be expressed schematically as $ G_{\rm eff}\sim \phi^{-1}$.

Within the framework of Conformal Cyclic Cosmology (CCC) \cite{Penrose1,Penrose2}, the Universe is described as a succession of aeons, each representing a complete cosmological epoch. An aeon originates in a Big-Bang state and evolves through an extended phase of expansion, ultimately approaching a de Sitter-like regime. Its remote future is represented by a spacelike conformal boundary, which provides the geometric endpoint of the aeon \cite{Penrose3}. Owing to the conformal invariance of the relevant asymptotic geometry, this future boundary can be identified with the Big-Bang surface of the subsequent aeon, thereby providing a cyclic description of cosmic evolution. A central element of this picture is the evolution of matter and gravitationally bound structures over the lifetime of an aeon. As the Universe expands, its temperature and matter density decrease, while an increasing fraction of matter may become incorporated into black holes. On sufficiently long timescales, black holes can lose their mass through Hawking radiation once the surrounding radiation temperature becomes lower than their effective Hawking temperature \cite{Hawking1}. In the asymptotic future, the complete evaporation of black holes leaves an effectively massless, radiation-dominated state. Under such conditions, the distinction between physical scales becomes irrelevant from the conformal point of view, allowing the remote future of one aeon to be conformally related to the Big-Bang state of the next one \cite{Penrose4}. This conformal correspondence provides the basis for the cyclic interpretation of the cosmological evolution in CCC.

A further observational motivation for considering Conformal Cyclic Cosmology (CCC) comes from unusual structures potentially present in the cosmic microwave background (CMB). An et al. \cite{An} identified anomalous circular features in Planck and WMAP maps, exhibiting a temperature rise toward their centers and amplitudes more than an order of magnitude larger than those of typical CMB fluctuations. They proposed that these features could be associated with Hawking points, hypothetical signatures of supermassive black holes evaporating in a previous aeon, as envisaged in CCC. According to the authors, such structures lack an obvious conventional explanation within the standard inflationary paradigm. Nevertheless, their identification as Hawking points remains tentative and requires further investigation. These findings therefore provide a possible, though not conclusive, observational motivation for exploring signatures of transitions between successive aeons.

The Little Red Dots (LRDs) are a population of compact, red sources discovered by the James Webb Space Telescope (JWST) at high redshifts with a number density peaking around $z\sim5$ \cite{Matthee2024,Kocevski2025}. Their physical nature remains debated. Proposed explanations include young, rapidly accreting supermassive black holes embedded in dense gas and dust, as well as compact, intensely star-forming systems \cite{PacucciLoeb2025,Kokorev}. Recent spectroscopic observations have strengthened the case for an active galactic nucleus origin in at least a substantial fraction of LRDs, although their unusual spectral properties remain incompletely understood \cite{Rusakov2026}.

The paper is organized as follows. Section \ref{sec2} introduces the Brans--Dicke model with a scalar-field potential. Section \ref{sec3} presents the power-law ansatz for the scalar field and the corresponding reconstruction of the scalar-field potential. Section \ref{sec4} discusses a heuristic physical interpretation of the $\phi\to\infty$ singularity, focusing on the possibility of accelerated black-hole evaporation as the effective gravitational coupling approaches zero. Section \ref{sec5} explores a phenomenological scenario for the transition between successive cosmological cycles and discusses its possible implications. Finally, Section \ref{sec6} summarizes the main results and outlines directions for further investigation.

\section{Brans--Dicke Model}
\label{sec2}

The Brans--Dicke (BD) theory is a scalar--tensor theory of gravity in which the gravitational interaction is mediated by both the tensor gravitational field and an additional scalar field $\phi$ \cite{BransDicke,Capozziello}. In the Jordan frame, the scalar field determines the effective gravitational coupling, which can be schematically expressed as
\be
G_{\rm eff} \sim \frac{1}{\phi}~.
\ee
The action of the Brans--Dicke theory is given by
\bea
\label{BDAction}
S= \frac{1}{16 \pi} \int d^4 x \sqrt{-g}
\left(
\phi R
-\frac{\omega}{\phi}
\partial_{\mu}\phi\partial^{\mu}\phi
-V \left( \phi \right) \right)
+S_{\rm matter}~,
\eea
where $\omega$ is the dimensionless Brans--Dicke coupling parameter, $\phi$ is the scalar field, and $V(\phi)$ is its potential. In the following, we assume $\omega>-3/2$, which is the standard condition for avoiding a ghost-like scalar degree of freedom in the Brans--Dicke theory.

Varying the action in Eq.~(\ref{BDAction}) with respect to the metric $g^{\mu\nu}$ yields the field equations
\bea
\label{fieldBDEQ}
G_{\mu \nu}= \frac{8 \pi}{\phi} T_{\mu \nu}+ \frac{\omega}{\phi^2} \left( \partial_{\mu} \phi \partial_{\nu} \phi - \frac{1}{2} g_{\mu \nu} \partial_{\alpha} \phi \partial^{\alpha} \phi \right) + \frac{1}{\phi} \left( \nabla_{\mu} \nabla_{\nu} \phi - g_{\mu \nu} \Box \phi \right) - g_{\mu \nu} \frac{V \left( \phi \right)}{2 \phi} ~,
\eea
where $G_{\mu\nu}$ is the Einstein tensor and $T_{\mu\nu}$ is the energy--momentum tensor. Varying the action (\ref{BDAction}) with respect to the scalar field $\phi$ gives
\bea
\label{DAlembert}
\Box \phi=\frac{1}{3+2 \omega} \left( 8 \pi T + \phi \frac{d V \left( \phi \right)}{d \phi} -2 V \left( \phi \right) \right)~,
\eea
where $\Box \equiv \nabla_{\mu}\nabla^{\mu}$ is the covariant d'Alembertian and $T=g^{\mu\nu}T_{\mu\nu}$ is the trace of the energy--momentum tensor. In the appropriate limit $\omega\rightarrow\infty$, with the scalar field approaching a constant value, the theory recovers General Relativity. The scalar field therefore introduces an additional dynamical degree of freedom that can modify the effective strength of the gravitational interaction.

In the following, we consider a spatially flat cosmological model, $k=0$. The corresponding FLRW line element is
\bea
\label{FriedmannMetric}
ds^2
=-dt^2
+
a^2(t)
\left[
dr^2
+
r^2
\left(
d\theta^2+\sin^2\theta\,d\varphi^2
\right)
\right]~,
\eea
where $a(t)$ is the scale factor. Substituting this metric into Eqs.~(\ref{fieldBDEQ}) and (\ref{DAlembert}) yields the equations governing the energy density and pressure:
\bea
\label{BDdensity}
\rho = \frac{\phi }{8 \pi} \left( \frac{3 \dot{a}^2 }{ a^2 } + \frac{3 \dot{a}  \dot{\phi} } {a  \phi }-\frac{\omega \dot{\phi}^2 }{2 \phi^2 } - \frac{V \left( \phi \right)}{2 \phi} \right)~,
\eea
\bea
\label{BDpressure}
p = -\frac{\phi }{8 \pi} \left(  \frac{2 \ddot{a}}{a}+ \frac{\dot{a}^2}{a^2} +\frac{\omega}{2} \frac{\dot{\phi}^2}{\phi^2} + \frac{2 \dot{a}\dot{\phi}}{a\phi} +\frac{\ddot{\phi}}{\phi} - \frac{V \left( \phi \right)}{2 \phi} \right)~,
\eea
together with the scalar-field equation, Eq.~(\ref{DAlembert}), which takes the form
\bea
\label{DalembertFM}
-\ddot{\phi}- \frac{3 \dot{a} \dot{\phi}}{a}= \frac{1}{3+2 \omega} \left[ 8 \pi \left( - \rho + 3p \right) + \phi \frac{d V \left( \phi \right)}{d \phi} -2 V \left( \phi \right) \right] ~.
\eea
Here, we have assumed a homogeneous scalar field, $\phi=\phi(t)$, and used the corresponding expression for the trace of the energy--momentum tensor, $T=-\rho+3p$.

\section{Power-Law Ansatz and the Scalar-Field Potential}
\label{sec3}

For the Brans--Dicke model introduced in Sec.~\ref{sec2}, we adopt a power-law scale factor of the form
\bea
\label{ScaleFactor}
a(t)=a_0 t^m~,
\eea
where $a_0$ and $m$ are constants. We also assume the following power-law form for the scalar field:
\bea
\label{ScalarField}
\phi(t)=\phi_0 (t_s-t)^n~,
\eea
where $\phi_0$ is a constant, $t_s$ denotes the time at which the future singularity occurs, and $n$ is a constant exponent. The inverse relation can be written as
\bea
\label{InverseScalarTime}
t(\phi)=t_s-\left(\frac{\phi}{\phi_0}\right)^{1/n}~.
\eea
To determine the derivative of the potential, $dV(\phi)/d\phi$, we substitute Eqs.~(\ref{BDdensity}), (\ref{BDpressure}) and (\ref{ScalarField}) into the scalar-field equation, Eq.~(\ref{DalembertFM}), and use Eq.~(\ref{InverseScalarTime}) to express the result in terms of $\phi$. This yields
\bea
\label{PotencialDerivative}
\frac{d V (\phi)}{d \phi} =
\frac{12 m^2 + 6 m \left[ n \omega \left[ t_s \left( \frac{\phi}{\phi_0} \right)^{-\frac{1}{n}} - 1 \right] - 1 \right]}
{\left[ t_s - \left( \frac{\phi}{\phi_0} \right)^{\frac{1}{n}} \right]^2 }
- (n-2)n\omega \left( \frac{\phi}{\phi_0} \right)^{-\frac{2}{n}}~.
\eea
The potential can then be obtained by integrating Eq.~(\ref{PotencialDerivative}). Since the resulting integral depends on the exponent $n$, we focus on a choice of $n$ for which an analytical expression can be obtained. In particular, we are interested in the case $n<0$, for which the effective gravitational coupling satisfies $G_{\rm eff}\sim 1/\phi\rightarrow 0$ as $t\rightarrow t_s^-$.

For $n=-1$, the integral can be evaluated analytically. Choosing $m=1/2$ and $\omega=-1$, we obtain
\bea
\label{PotencialVPhi}
V(\phi)=V_0+\frac{3\phi}{t_s^2}
+\frac{3\phi^2}{2t_s\phi_0}
+\frac{\phi^3}{\phi_0^2}
+\frac{3\phi_0}{t_s^3}
\ln\left|1-\frac{t_s\phi}{\phi_0}\right|~,
\eea
where $V_0$ is an integration constant.

For the chosen parameters, the limit $t\rightarrow 0^+$ corresponds to an initial cosmological singularity. In this limit, the scale factor vanishes, $a(t)\rightarrow 0$, while the energy density and pressure diverge, $\rho\rightarrow\infty$ and $p\rightarrow\infty$. The scalar field, however, approaches the finite value $\phi\rightarrow\phi_0/t_s$. The potential becomes singular because the argument of the logarithm in Eq.~(\ref{PotencialVPhi}) approaches zero, such that $|V(\phi)|\rightarrow\infty$.

In the limit $t\rightarrow t_s^-$, the scale factor remains finite,
$a(t_s)=a_0\sqrt{t_s}$, whereas the energy density and the absolute value of the pressure diverge. More specifically, for $\phi_0>0$ and $t_s>0$, the leading asymptotic behavior is
\bea
\rho(t)\sim \frac{3\phi_0}{32\pi t_s}\frac{1}{(t_s-t)^2},
\qquad
p(t)\sim -\frac{\phi_0}{8\pi}\frac{1}{(t_s-t)^3}~.
\eea
This behavior is characteristic of a Type III (finite-scale-factor) cosmological singularity: the scale factor remains finite, while the energy density and pressure diverge. In the present model, the divergence is associated with the scalar field, which satisfies $\phi(t)\rightarrow\infty$ as $t\rightarrow t_s^-$, rather than with a divergence of the scale factor.

In a more realistic cosmological evolution, matter may eventually dominate over radiation, corresponding approximately to $m=2/3$ in a standard matter-dominated regime. The present analysis, however, employs a single power-law ansatz with $m=1/2$. A transition between these regimes would require a more general treatment; its effect on the detailed asymptotic behavior should therefore be examined separately.

\section{Physical interpretation of the $\phi\rightarrow\infty$ singularity}
\label{sec4}

The scalar-field singularity is particularly important for the physical interpretation of the limit $\phi\rightarrow\infty$, which, within the Brans--Dicke framework considered here, corresponds to
\bea
G_{\rm eff}\rightarrow0~.
\eea
Thus, the approach to the singularity is accompanied by a progressive suppression of the effective gravitational interaction. As a consequence, gravitationally bound structures may become increasingly susceptible to disruption as the singularity is approached. The resulting weakening of gravitational binding could facilitate the disruption of structures on progressively larger scales and the redistribution of their energy among less gravitationally bound degrees of freedom.

This behavior can also be considered in the context of black holes. As a phenomenological estimate, we employ the standard Schwarzschild expression for the gravitational radius, written in terms of the effective gravitational coupling:
\bea
r_{\rm S}(t)=\frac{2G_{\rm eff}(t)M}{c^2}~,
\eea
where $M$ denotes the black-hole mass. If $G_{\rm eff}(t)=G_{\rm ref}\phi_{\rm ref}/\phi(t)$, with $\phi_{\rm ref}$ and $G_{\rm ref}$ denoting reference values of the scalar field and the effective gravitational coupling, respectively, then
\bea
r_{\rm S}(t)=\frac{2G_{\rm ref}M}{c^2}
\frac{\phi_{\rm ref}}{\phi(t)}~.
\eea
For a fixed mass $M$, this expression implies
\bea
r_{\rm S}(t)\rightarrow0
\qquad
\text{as}
\qquad
\phi(t)\rightarrow\infty~.
\eea
Within this simplified interpretation, the characteristic gravitational radius associated with a black hole decreases as the singularity is approached. This suggests a possible mechanism by which matter initially confined within compact gravitational structures could progressively lose its gravitational confinement. One may envisage scenarios in which black holes form during an earlier stage of the evolution, while the subsequent weakening of the effective gravitational interaction reduces the characteristic gravitational length scale associated with these objects. If the black holes subsequently evaporate, the released energy could contribute to a radiation-dominated state near the singularity.

For the present analysis, the standard Schwarzschild radius is used only as a phenomenological estimate. In Brans--Dicke gravity, however, black-hole solutions can differ substantially from their general-relativistic counterparts. Various static and dynamical solutions have been discussed in the literature and, depending on the assumptions imposed on the scalar field and the spacetime geometry, some solutions describe naked singularities or wormhole-like configurations rather than conventional black-hole horizons \cite{Faraoni}. Consequently, the behavior of $r_{\rm S}$ considered above should not be interpreted as a general theorem concerning black-hole horizons in Brans--Dicke theory, but rather as an indication of how a vanishing effective gravitational coupling might affect compact gravitational structures.

It should nevertheless be stressed that this interpretation extends beyond the regime in which classical gravitational theory can be regarded as a complete description. In particular, the behavior of black-hole horizons and the validity of the standard Hawking evaporation picture when $G_{\rm eff}$ varies dynamically and approaches zero cannot be established from the conventional semiclassical treatment alone. A complete description of this limiting regime would ultimately require a consistent theory of quantum gravity. The proposed conversion of gravitationally bound matter into radiation should therefore be regarded as a possible physical interpretation of the asymptotic behavior of the model, rather than as a direct prediction of the classical Brans--Dicke field equations.

A closely related mechanism was previously considered in the context of disformal gravity with dynamical fundamental constants. The results obtained here within the Brans--Dicke framework exhibit a qualitatively similar behavior to that found in the disformal model. In both approaches, the weakening of the effective gravitational interaction provides a possible mechanism for the progressive disruption of gravitationally bound structures and the conversion of their remaining energy into radiation \cite{Marosek3}.

\section{Crossover into the Next Cycle}
\label{sec5}

At present, we have not identified a specific physical mechanism that would allow a transition between successive branches of the scalar field. Nevertheless, we may phenomenologically assume that, for $t_{s(k)}<t<t_{s(k+1)}$, the scalar field evolves according to
\bea
\label{ScalarFieldNC}
\phi(t)=\phi_{k+1}\left(t_{s(k+1)}-t\right)^n,
\eea
where $\phi_{k+1}$ is a constant that need not coincide with the constant $\phi_k$ characterizing the preceding branch, and $t_{s(k+1)}>t_{s(k)}$ denotes the time at which the subsequent scalar-field singularity occurs. In contrast to the preceding branch, the scalar field is assumed to take a finite value immediately after the transition at $t=t_{s(k)}$. In particular, if the effective gravitational coupling satisfies $G_{\rm eff}\propto 1/\phi$, then $G_{\rm eff}\to 0$ as $t\to t_{s(k)}^-$, whereas it approaches a finite value as $t\to t_{s(k)}^+$, provided that $\phi_{k+1}>0$ and the parameters are regular.

For the cases considered here, we have not found a scalar-field ansatz that yields solutions on both sides of the singularity while preserving a common scalar-field potential. Equation~(\ref{ScalarFieldNC}) should therefore be understood as a phenomenological prescription rather than as a transition derived from the classical Brans--Dicke field equations. The transition between successive branches remains an open problem in the present model. A consistent description of this regime may require a more fundamental treatment of the scalar and gravitational degrees of freedom, potentially including quantum effects beyond the classical framework employed here.

For the scale factor, we retain the functional form adopted in the preceding branch,
\bea
\label{ScaleFactorNC}
a(t)=a_0t^m,\qquad t>t_{s(k)}.
\eea
For the scalar-field ansatz given above, the potential must be reconstructed anew. For the same parameter values as in the preceding branch, namely $\omega=-1$, $m=1/2$, and $n=-1$, we obtain
\bea
\label{PotencialDerivativeNC}
\frac{dV(\phi)}{d\phi}
=
\frac{3t_{s(k+1)}\phi^3}
{\left(t_{s(k+1)}\phi-\phi_{k+1}\right)\phi_{k+1}^2},
\eea
and hence
\bea
\label{PotencialNC}
V(\phi)=V_1
+\frac{3\phi}{t_{s(k+1)}^2}
+\frac{3\phi^2}{2t_{s(k+1)}\phi_{k+1}}
+\frac{\phi^3}{\phi_{k+1}^2}
+\frac{3\phi_{k+1}}{t_{s(k+1)}^3}
\ln\left|1-\frac{t_{s(k+1)}\phi}{\phi_{k+1}}\right|,
\eea
where $V_1$ is an integration constant. This expression generally differs from the potential reconstructed for the preceding branch. Thus, the two branches do not, in the construction considered here, arise from a single common potential while maintaining the chosen scalar-field ansatz and a positive scalar field on both branches.

For the parameter values specified above, the asymptotic behaviour of the energy density and pressure as $t\to t_{s(k+1)}^{-}$ is given by
\bea
\rho(t)&\sim&
\frac{3\phi_{k+1}}{32\pi t_{s(k+1)}}
\frac{1}{(t_{s(k+1)}-t)^2},
\\
p(t)&\sim&
-\frac{\phi_{k+1}}{8\pi}
\frac{1}{(t_{s(k+1)}-t)^3}.
\eea
For $\phi_{k+1}>0$ and $t_{s(k+1)}>0$, the energy density therefore diverges to positive infinity, $\rho\to+\infty$, whereas the pressure diverges to negative infinity, $p\to-\infty$. At the same time, the scalar field and the potential diverge, $\phi\to+\infty$ and $V(\phi)\to+\infty$, while the scale factor approaches the finite value
$a(t_{s(k+1)})=a_0t_{s(k+1)}^{1/2}$.
The terminal singularity of this branch is thus characterized by divergent energy density and pressure despite the finite value of the scale factor.

At the beginning of the new branch, $t\to t_{s(k)}^+$, the scalar field instead approaches the finite value
$\phi(t_{s(k)}^+)=\phi_{k+1}(t_{s(k+1)}-t_{s(k)})^{-1}$.
For regular choices of the model parameters, the energy density, pressure, and potential are also finite there. The proposed construction therefore connects a branch ending in a scalar-field singularity to a new branch that initially has finite scalar-field and matter variables.

The broader cosmological interpretation considered here is necessarily speculative. In this scenario, the Universe begins with a Big Bang and subsequently passes through a radiation-dominated era. As the Universe expands and cools, matter structures form and the matter content eventually becomes dynamically dominant. Gravitational collapse then leads to the formation of black holes, and a substantial fraction of the matter may ultimately be accreted by them.

As the subsequent scalar-field singularity is approached, the effective gravitational coupling tends to zero, $G_{\rm eff}\to0$. We may speculate that the changing gravitational regime could be associated with a rapid release of energy from black holes, particularly those residing in galactic nuclei, in the form of radiation. If a sufficiently dense radiation field were produced, it could, in principle, generate particles and antiparticles and thereby replenish the matter content of the Universe. In a further extension of this picture, the resulting matter and radiation could contribute to the formation of new structures, potentially including structures in galactic nuclei. This would provide a possible qualitative mechanism for the recurrent appearance of galactic structures in successive cycles, although its physical viability remains to be established.

\section{Conclusions}
\label{sec6}

In this work, we have investigated a Brans--Dicke cosmological model with a scalar-field potential and obtained an analytical solution for a power-law ansatz for the scalar field. The terminal singularity found in this model is driven by the evolution of the scalar field. In terms of the energy density $\rho$, pressure $p$, and scale factor $a(t)$, it exhibits the characteristic features of a Type III singularity, also referred to as a finite-scale-factor singularity (FSFS): the energy density and pressure diverge, whereas the scale factor remains finite. At the singularity, the scalar field diverges, $\phi\to\infty$, and the effective gravitational coupling consequently vanishes, $G_{\rm eff}\to0$.

Within a heuristic interpretation of the model, we have considered the possibility that the evolution of the effective gravitational coupling could influence the fate of black holes. In particular, the Schwarzschild radius is proportional to the gravitational constant for a black hole of fixed mass. This observation motivates the hypothesis that a sufficiently rapid decrease in $G_{\rm eff}$ could be associated with accelerated black-hole evaporation. In our model, the rate at which the effective gravitational coupling decreases near the singularity depends on the scalar-field exponent $n$. For negative values of $n$ with small absolute magnitude, the scalar field may evolve relatively slowly over much of the cosmological evolution before increasing rapidly as the singularity is approached. The corresponding decrease in $G_{\rm eff}$ could therefore become significant over a comparatively short interval. Whether this behaviour can actually induce rapid black-hole evaporation, however, requires a separate dynamical analysis and is not established by the present solution alone.

We have also phenomenologically postulated a transition through the scalar-field singularity, connecting one cosmological branch to a subsequent branch. We have not found a solution within the ansatz considered here that provides both a smooth transition between the branches and a common scalar-field potential on either side of the singularity. Moreover, the present model does not specify a physical mechanism responsible for this transition. The problem therefore remains open and may require a more fundamental theoretical framework, potentially involving physics beyond the classical Brans--Dicke theory.

Despite these limitations, the phenomenological transition allows us to consider a possible post-singularity scenario. In this picture, the effective gravitational coupling becomes finite again after the transition. We further hypothesize that, during the transition, black holes could release a substantial fraction of their mass in the form of radiation. If this radiation were sufficiently dense, it might give rise to renewed particle--antiparticle production and, subsequently, to the formation of matter. With a nonzero effective gravitational coupling restored, the resulting matter could potentially participate in the formation of new gravitational structures, including structures in galactic nuclei. This scenario offers a speculative picture of how some structures might reappear in successive cosmological cycles, but its physical consistency and dynamical realization remain to be demonstrated.

The scenario raises the speculative possibility that some of the objects observed by the James Webb Space Telescope and referred to as Little Red Dots could be related to processes occurring near a transition between cosmological cycles. While interpretations involving accreting black holes have been proposed for these objects, one might alternatively ask whether, in the present scenario, some of their observed emission could arise from matter production associated with dense radiation generated during a hypothetical episode of black-hole evaporation. In a still more speculative interpretation, such objects might retain information about a preceding cosmological cycle. There is currently no observational evidence supporting this proposal, and it should be regarded as a conjecture rather than a prediction of the model.

The results presented here should therefore be viewed as an exploratory investigation of the possible cosmological implications of a scalar-field singularity in Brans--Dicke theory. Although the analytical solution provides a concrete setting in which the effective gravitational coupling vanishes, the proposed black-hole evaporation scenario, the transition between successive branches, and the possible observational connections remain open questions for future investigation.\\

\textbf{Acknowledgments} The authors would like to acknowledge the use of OpenAI's ChatGPT for assistance with language editing, paraphrasing, and improving the clarity and organization of the manuscript. All scientific content, calculations, interpretations, and conclusions were developed and verified by the authors.


\begin{thebibliography}{99}
%
%

\bibitem{HawkingElis}
Hawking, S. W. and Ellis G. F. R., The Large Scale Structure of Space-Time, \emph{Cambridge University Press}, (1973).

\bibitem{Riess}
 Riess A. G. et al., Observational evidence from supernovae for an acceler-ating universe and a cosmological constant, \emph{Astronomical J.} {\bf 116}, 1009 (1998).

\bibitem{Perlmutter}
Perlmutter S. et al., Measurements of $\Omega$ and $\Lambda$ from 42 High-Redshift Supernovae \emph{ApJ} {\bf 517}, 565 (1999).

\bibitem{Tonry}
Tonry J.L. et al., Cosmological Results from High-z Supernovae \emph{The Astrophysical Journal} {\bf 594}, 1 (2003).

\bibitem{Knop}
Knop R. A. et al., New Constraints on $ \Omega_{M} $, $ \Omega_{\Lambda} $, and w from an Independent Set of 11 High-Redshift Supernovae Observed with the Hubble Space Telescope \emph{The Astrophysical Journal} {\bf 598}, 102 (2003).

\bibitem{Tegmark}
Tegmark M. et al., Cosmological parameters from SDSS and WMAP \emph{Phys. Rev. D} {\bf 69}, 103501 (2004).

\bibitem{Hinshaw}
Hinshaw G. et al., Nine-Year Wilkinson Microwave Anisotropy Probe (WMAP) Observations: Cosmological Parameter Results, \emph{ApJS} {\bf 208}, 19 (2013).

\bibitem{Haro}
Haro J., Nojiri S., Odintsov S. D., Oikonomou V. K., Pan S. Finite-time cosmological singularities and the possible fate of the Universe, \emph{Phys. Rep.}  {\bf 1034}, 1-114 (2023).

\bibitem{Dabrowski1}
D\c{a}browski M. P., Stachowiak T., Szydłowski M. Phantom Friedmann Cosmologies and Higher-Order Characteristics of Expansion, \emph{Phys. Rev. D} {\bf 68}, 103519 (2003).

\bibitem{Caldwell}
Caldwell R. R. A phantom menace? Cosmological consequences of a dark energy component with super-negative equation of state, \emph{Phys. Lett. B} {\bf 545}, 23 (2002).

\bibitem{Barrow}
Barrow J. D., and Tsagas C. G., New isotropic and anisotropic sudden singularities, \emph{Classical and Quantum Gravity} {\bf 22}, 9 (2005).

\bibitem{Barrow1}
Barrow J. D., Sudden future singularities \emph{Class. Quantum Grav.} {\bf 21} 79 (2004).

\bibitem{Odintsov1}
S. Nojiri and S. D.Odintsov, The final state and thermodynamics of dark energy universe, \emph{Phys.Rev. D} {\bf{70}}, 103522 (2004).

\bibitem{Dabrowski2}
M.P. D\c{a}browski and T. Denkiewicz Exotic-singularity-driven dark energy, \emph{AIP Conference Proceedings} {\bf 1241}, 561 (2010).

\bibitem{Odintsov2}
Nojiri, S., Odintsov S. D. and Tsujikawa S. Properties of singularities in (phantom) dark energy universe, \emph{Phys.Rev. D} {\bf 71}, 063004 (2005).

\bibitem{Dabrowski3}
D\c{a}browski M. P., Denkiewicz T. Barotropic index w-singularities in cosmology, \emph{Phys. Rev. D} {\bf 79}, 063521 (2009).

\bibitem{Fernandez}
Fern\'andez-Jambrina L., $w$-cosmological singularities, \emph{Phys. Rev. D} {\bf 82}, 124004 (2010).

\bibitem{Frampton}
Frampton P. H., Ludwick K. J. and Scherrer R. J. The little rip, \emph{Phys. Rev. D} {\bf 84} 063003 (2011).

\bibitem{Frampton1}
Frampton P. H., Ludwick K. J. and Scherrer R. J. Pseudo-rip: Cosmological models intermediate between the cosmological constant and the little rip, \emph{Phys. Rev. D} {\bf 85} 083001 (2012).

\bibitem{Curiel}
Curiel E., A Primer on Energy Conditions. \emph{Towards a Theory of Spacetime Theories. Einstein Studies}, vol {\bf 13}. Birkhäuser, New York, NY (2017).

\bibitem{Tipler}
Tipler F., Singularities in conformally flat spacetimes, \emph{Phys. Lett. A} {\bf 64}, 8 (1977).

\bibitem{Królak}
Królak A., Towards the proof of the cosmic censorship hypothesis, \emph{Class. Quantum Grav.} {\bf 3}, 267 (1986).

\bibitem{Houndjo}
Houndjo  M.J.S., Batista C.E.M., Campos J.P., and Piattella O.F., Finite-time singularities in f$(R, T)$ gravity and the effect of conformal anomaly, \emph{Canadian Journal of Physics} {\bf 91}, 7 (2013).

\bibitem{Marosek}
D\c{a}browski M. P. and Marosek K., Regularizing cosmological singularities by varying physical constants, \emph{JCAP} {\bf 02}, 012 (2013).

\bibitem{Nojiri}
Nojiri S. and Odintsov S. D., The dark universe future and singularities: The account of thermal and quantum effects, \emph{Physics of the Dark Universe} Volume {\bf 30}, 100695 (2020).

\bibitem{Elizalde}
Elizalde E., Yurov A.V. and Timoshkin, A.V., Singularity softening and avoidance by the action of thermal radiation in a generalized entropic cosmology, \emph{Eur. Phys. J. C} {\bf 85}, 1375 (2025).

\bibitem{BransDicke}
Brans C. and Dicke R.H., Mach's Principle and a Relativistic Theory of Gravitation, \emph{Phys. Rev.} {\bf 124}, 925 (1961).

\bibitem{Penrose1}
Penrose R., Cycles of Time: An Extraordinary New View of the Universe, \emph{Bodley Head, London}, (ISBN 9780224080361).

\bibitem{Penrose2}
Penrose R., Before the big bang: an outrageous new perspective and its implications for particle physics, In EPAC 2006 Proc., Edinburgh, Scotland, pp.2759-62 ed. C.R.Prior., (2006).

\bibitem{Penrose3}
Penrose R., Republication of: Conformal treatment of infinity, \emph{Gen Relativ Gravit} {\bf 43}, 901–922 (2011).

\bibitem{Hawking1}
Hawking S.W., Particle creation by black holes, \emph{Comm. Math. Phys.} {\bf 43}, 199-220, (1975).

\bibitem{Penrose4}
Penrose R., The basic ideas of conformal cyclic cosmology, \emph{AIP Conf. Proc.} {\bf1446}, 233–243 (2012), https://doi.org/10.1063/1.4727997

\bibitem{An}
An D., Meissner K. A., Nurowski P., Penrose R., Apparent evidence for Hawking points in the CMB Sky, \emph{Monthly Notices of the Royal Astronomical Society}, Volume {\bf 495}, 3: 3403–3408, (2020).

\bibitem{Matthee2024}
Matthee J. et al., Little Red Dots: An Abundant Population of Faint Active Galactic Nuclei at $z\sim 5$ Revealed by the EIGER and FRESCO JWST Surveys, \emph{The Astrophysical Journal} {\bf 963} 129 (2024).

\bibitem{Kocevski2025}
Kocevski D. D. at al., The Rise of Faint, Red Active Galactic Nuclei at $z > 4$: A Sample of Little Red Dots in the JWST Extragalactic Legacy Fields, \emph{The Astrophysical Journal} {\bf 986} 126 (2024).

\bibitem{PacucciLoeb2025}
Pacucci F. and Loeba A. Cosmic Outliers: Low-spin Halos Explain the Abundance, Compactness, and Redshift Evolution of the Little Red Dots, \emph{The Astrophysical Journal Letters} {\bf 989} L19 (2025).

\bibitem{Kokorev}
Kokorev V. et al, The Deepest GLIMPSE of a Dense Gas Cocoon Enshrouding a Little Red Dot, \emph{ApJ} {\bf 1004} 153 (2026).

\bibitem{Rusakov2026}
Rusakov V. at al. Little red dots as young supermassive black holes in dense ionized cocoons \emph{Nature} {\bf 649} 574-579 (2026).

\bibitem{Capozziello}
Capozziello S. and De Laurentis M. Extended Theories of Gravity, \emph{Physics Reports} {\bf 509}  4–5, Pages 167-321 (2011).

\bibitem{Faraoni}
Faraoni V., Hammad F., and Belknap-Keet S. D., Revisiting the Brans solutions of scalar-tensor gravity, \emph{Phys. Rev. D} {\bf 94}, 104019 (2016).

\bibitem{Marosek3}
Marosek K. and Balcerzak A., Conformal Cyclic Cosmology from Varying Fundamental Constants, https://arxiv.org/abs/2608.27014



\end{thebibliography}
\end{document}